# The View Update Problem Revisited

Enrico Franconi    Paolo Guagliardo

Free University of Bozen-Bolzano
<surname>@inf.unibz.it

## Abstract

In this paper, we revisit the view update problem in a relational setting and propose a framework based on the notion of determinacy under constraints. Within such a framework, we characterise when a view mapping is invertible, establishing that this is the case precisely when each database symbol has an exact rewriting in terms of the view symbols under the given constraints, and we provide a general effective criterion to understand whether the changes introduced by a view update can be propagated to the underlying database relations in a unique and unambiguous way.

Afterwards, we show how determinacy under constraints can be checked, and rewritings effectively found, in three different relevant scenarios in the absence of view constraints. First, we settle the long-standing open issue of how to solve the view update problem in a multi-relational database with views that are projections of joins of relations, and we do so in a more general setting where views are defined by arbitrary conjunctive queries and database constraints are stratified embedded dependencies. Next, we study a setting based on horizontal decompositions of a single database relation, where views are defined by selections on possibly interpreted attributes (e.g., arithmetic comparisons) in the presence of domain constraints over the database schema. Lastly, we look into another multi-relational database setting, where views are defined in an expressive *Type* Relational Algebra based on the $n$-ary Description Logic $\mathcal{DLR}$ and database constraints are inclusions of expressions in that algebra.

## 1 Introduction

Updating a database by means of a set of views is a challenging task, known as the *view update problem*, that requires updates of the views to be translated into suitable updates of the underlying database that consistently propagate the changes of the views onto the base relations over which the views are defined. For an overview by examples of the difficulties arising in this context refer to the survey [16]. In a nutshell, solving the view update problem essentially boils down to (1) understanding whether the views are in fact "updatable", which amounts to checking whether the mapping from the database relations to the views is invertible, as we shall see; and (2) establishing whether a view update is "translatable", that is, whether it has a translation.

A general and precise understanding of the view update problem is due to Bancilhon and Spyratos [1], who provide an elegant solution to it within an abstract functional framework. However, no effective methods for checking the translatability of updates and computing their translations are actually provided in [1], where it is asserted that "computational algorithms (if they exist) must be sought in specific problems". Indeed, the view update problem is formalised at a high level of abstraction in [1], where views and updates are arbitrary functions, of which no constructive characterisation is given, as one might not even be possible.

To the best of our knowledge, [3] is the only comprehensive work in which the view updated problem is investigated in the context of relational databases. However, the setting considered there is very restricted, as it is limited to only two views defined by projections over a single database relation. This was recently generalised in [12] to an arbitrary number of views defined by acyclic projections, but still over a database consisting of a single relation.

**Contribution.** With this work we contribute the following:

- A general view update framework, based on the notion of determinacy, that constructively revisits [1] in a relational setting with constraints.
- A general criterion, applicable in all the cases in which the inverse of a view mapping can be effectively computed, for checking whether a view update can be propagated to the underlying database in a unique and unambiguous way.
- The analysis of three different relevant settings where the problem of checking whether a view mapping is invertible is decidable, and the inverse can be effectively computed. The first analysed scenario provides a general solution to a long-standing open issue pointed out in [3], namely how to solve the view update problem in a multi-relational database with views that are projections of joins of relations.

**Outline.** The rest of the article is organised as follows: we start by introducing notation, basic definitions and few other relevant notions in Section 2; we continue by presenting our framework and our main res-ults concerning invertibility of views and translatability of view updates in Section 3; we then discuss the three different scenarios for which we can effectively solve the view update problem, namely one setting where views are defined by conjunctive queries over a multi-relational database with stratified embedded dependencies (Section 4), one where views are defined by selections over a single-relation database with domain constraints on interpreted attributes (Section 5), and one other setting with a multi-relational database in which views and constraints are expressed in an expressive fragment of relational algebra corresponding to an $n$-ary description logic (Section 6); finally, we briefly discuss related work in Section 7 and conclude in Section 8 by pointing out possible future research directions.

For the sake of clarity in the presentation of our results, the proofs are not included in the main body of the article, but they can be found in Appendix A, along with additional definitions and notation that might be eventually needed.

## 2 Preliminaries

We begin with some notation and basic definitions. Mostly, we use standard terminology from database theory.

**Basics.** An *n-ary relation* on a set $A$ is a set of $n$-tuples of elements of $A$. Tuples are denoted with an overline (e.g., $\overline{t}$). A *signature* (or *schema*) is a finite set $\mathcal{S}$ of relation symbols, where each $S \in \mathcal{S}$ has a non-negative arity denoted by $|S|$. Let **dom** be an arbitrary (possibly infinite) set of domain values. An *instance* $I$ over a signature $\mathcal{S}$ maps each relation symbol $S$ in $\mathcal{S}$ to a relation $S^I$ on **dom** of appropriate arity, called the *extension* of $S$ under $I$.[1] The set of elements of **dom** that occur in an instance $I$ is called the *active domain* of $I$ and is denoted by $\mathbf{adom}(I)$. An instance is finite when its active domain is, and we always assume instances to be finite unless otherwise specified. The *disjoint union* $I \uplus J$ of any two instances $I$ and $J$ over disjoint signatures $\mathcal{S}$ and $\mathcal{T}$, respectively, is the instance with active domain $\mathbf{adom}(I) \cup \mathbf{adom}(J)$ associating each $S \in \mathcal{S} \cup \mathcal{T}$ with $S^I$ if $S \in \mathcal{S}$ and with $S^J$ otherwise. Given an instance $I$ over a signature $\mathcal{S}'$, we denote by $\mathcal{S}^I$ the restriction of $I$ to $\mathcal{S} \subseteq \mathcal{S}'$. We indicate as a sub-script the signature over which an instance is, e.g., $I_\mathcal{S}$ is an instance over $\mathcal{S}$.

A *database schema* is a signature $\mathcal{R}$ of *database symbols* and a *database state* is an instance over $\mathcal{R}$. A *view schema* is a signature $\mathcal{V}$ of *view symbols* not occurring in $\mathcal{R}$ and a *view state* is an instance over $\mathcal{V}$. The set of all database states (resp., view states) is denoted by $\mathbf{R}$ (resp., $\mathbf{V}$). For a database state $I_\mathcal{R} \in \mathbf{R}$ and a view state $I_\mathcal{V} \in \mathbf{V}$, the instance $I_\mathcal{R} \uplus I_\mathcal{V}$ is called a *global state* over $\mathcal{R} \cup \mathcal{V}$.

---
[1] Sometimes, we use $S$ to denote informally both the relation symbol and the relation $S^I$ interpreting it.

**Constraints.** A *constraint* is a closed formula $\varphi$ in some domain-independent function-free fragment $\mathfrak{L}$ of first-order logic (FOL) over the signature $\mathcal{S}$ and constants **dom** under the standard name assumption. The sets of all the relation symbols and constants occurring in $\varphi$ are denoted by $\mathsf{sig}(\varphi)$ and $\mathsf{const}(\varphi)$, respectively. We extend $\mathsf{sig}(\cdot)$ and $\mathsf{const}(\cdot)$ to sets of constraints in the natural way. For a set of constraints $\Gamma$ and an instance $I$ over $\mathcal{S}$, we write $I \models \Gamma$ to indicate that the relational structure $\mathcal{I} = \langle \mathbf{adom}(I) \cup \mathsf{const}(\Gamma), I \rangle$ is a *model* of (or *satisfies*) $\Gamma$, that is, $\mathcal{I}$ makes every formula $\varphi$ in $\Gamma$ true. A set of constraints $\Gamma$ *entails* (or *logically implies*) a constraint $\varphi$ over $\mathsf{sig}(\Gamma)$, written $\Gamma \models \varphi$, if every finite instance satisfying $\Gamma$ also makes $\varphi$ true. Given two disjoint signatures $\mathcal{S}$ and $\mathcal{T}$ and a set of constraints $\Gamma$ over $\mathcal{S} \cup \mathcal{T}$, an instance $I_\mathcal{S}$ (resp., $I_\mathcal{T}$) is $\Gamma$-*consistent* if there exists an instance $I_\mathcal{T}$ (resp., $I_\mathcal{S}$) such that $I_\mathcal{S} \uplus I_\mathcal{T} \models \Gamma$. All sets of constraints we consider are finite.

Throughout the paper, we consider a satisfiable finite set $\Sigma$ of *global constraints* over $\mathcal{R} \cup \mathcal{V}$, partitioned into a set $\Sigma_\mathcal{R}$ of *database constraints* over $\mathcal{R}$, a set $\Sigma_\mathcal{V}$ of *view constraints* over $\mathcal{V}$, and a set $\Sigma_{\mathcal{R}\mathcal{V}}$ of *interschema constraints* over $\mathcal{R} \cup \mathcal{V}$ consisting of exactly one formula of the form $\forall \overline{x} \bigl( V(\overline{x}) \leftrightarrow \phi(\overline{x}) \bigr)$ for each $V \in \mathcal{V}$, where $\phi(\overline{x})$ is such that $\mathsf{sig}(\phi) \subseteq \mathcal{R}$ and is called a *definition* of $V$ in terms of $\mathcal{R}$. We denote the set of $\Sigma$-consistent database states (resp., view states) by $\mathbf{R}_\Sigma$ (resp., $\mathbf{V}_\Sigma$). If $\Sigma$ is understood from the context, we refer to $\Sigma$-consistent states also as *globally consistent states* or *states that are consistent with the global constraints*.

**Renamings.** A *renaming* over a signature $\mathcal{S}$ is a bijective function $\mathsf{ren}\colon \mathcal{S} \to \mathcal{S}'$, where $\mathcal{S}'$ is a signature disjoint with $\mathcal{S}$. We extend $\mathsf{ren}(\cdot)$ to instances and (sets of) constraints in the natural way. For example, the renaming $\mathsf{ren}(\varphi)$ of a constraint $\varphi$ is obtained from $\varphi$ by replacing every occurrence of each symbol $S \in \mathsf{sig}(\varphi)$ with $\mathsf{ren}(S)$. Clearly, for a set of constraints $\Gamma$ over $\mathcal{S}$ and an instance $I$ over $\mathsf{ren}(\mathcal{S})$, it is the case that $I \models \mathsf{ren}(\Gamma)$ iff $\mathsf{ren}^{-1}(I) \models \Gamma$.

We use the terms *function* and *mapping* interchangeably, and functions are assumed to be total unless specified otherwise. The *surjective restriction of* a function $f$ is the surjective mapping obtained from $f$ by restricting its codomain to its image. We use concatenation to indicate composition, e.g., $fg$ denotes the composition of $f$ with $g$.

**Views and determinacy.** Let $\mathcal{S}$ and $\mathcal{T}$ be two disjoint schemas, and let $\Gamma$ be a set of constraints over $\mathcal{S} \cup \mathcal{T}$.

*Definition 1* (View under constraints). A *view from $\mathcal{S}$ to $\mathcal{T}$ under $\Gamma$* is a function associating each $\Gamma$-consistent instance $I_\mathcal{S}$ with a $\Gamma$-consistent instance $I_\mathcal{T}$ such that $I_\mathcal{S} \uplus I_\mathcal{T} \models \Gamma$.

Observe that, as $\mathcal{S}$ and $\mathcal{T}$ are disjoint, every instance over $\mathcal{S} \cup \mathcal{T}$ satisfying $\Sigma$ has the form $I_\mathcal{S} \uplus I_\mathcal{T}$ where $I_\mathcal{S}$ and $I_\mathcal{T}$ are instances over $\mathcal{S}$ and $\mathcal{T}$, respectively.

*Definition 2 (Determinacy under constraints).* We say that $\mathcal{S}$ *determines* $\mathcal{T}$ *under* $\Gamma$ (and write $\mathcal{S} \twoheadrightarrow_\Gamma \mathcal{T}$) if, for every $I_\mathcal{S}$ and $I_\mathcal{T}, I'_\mathcal{T}$, it is the case that $I_\mathcal{T} = I'_\mathcal{T}$ whenever $I_\mathcal{S} \uplus I_\mathcal{T} \models \Gamma$ and $I_\mathcal{S} \uplus I'_\mathcal{T} \models \Gamma$.

In other words, models of $\Gamma$ that agree on the extension of the symbols in $\mathcal{S}$ also agree on the extension of the symbols in $\mathcal{T}$, which means that in every model of $\Gamma$ the latter functionally depends on former.

In general, there might exist more than one view mapping that satisfies a given set of constraints. An important connection between determinacy and views under constraints is that, not surprisingly, the view from $\mathcal{S}$ to $\mathcal{T}$ is unique exactly when the symbols in $\mathcal{T}$ are determined by the symbols in $\mathcal{S}$ under the given constraints.

THEOREM 1. *$\mathcal{S} \twoheadrightarrow_\Gamma \mathcal{T}$ if and only if there is exactly one view from $\mathcal{S}$ to $\mathcal{T}$ under $\Gamma$.*

In light of this, we write $\mathcal{S} \twoheadrightarrow_\Gamma^f \mathcal{T}$ to indicate that $\mathcal{S} \twoheadrightarrow_\Gamma \mathcal{T}$ and $f$ is the (one and only) view *induced by the constraints* $\Gamma$ from $\mathcal{S}$ to $\mathcal{T}$.

Observe that, under the assumptions we made about the global constraints $\Sigma$, it is always the case that $\mathcal{R} \twoheadrightarrow_\Sigma \mathcal{V}$ and there is exactly one view $f\colon \mathbf{R}_\Sigma \to \mathbf{V}_\Sigma$ from $\mathcal{R}$ to $\mathcal{V}$ under $\Sigma$, because the interschema constraints $\Sigma_{\mathcal{R}\mathcal{V}}$ explicitly define each view symbol in terms of the database symbols, while the database constraints $\Sigma_\mathcal{R}$ and the view constraints $\Sigma_\mathcal{V}$ together define the domain and codomain of the view $f$. In the rest of the paper, unless we specify otherwise, whenever we say "a view" we refer to the view from $\mathcal{R}$ to $\mathcal{V}$ induced by a set of global constraints $\Sigma$ partitioned as above.

## 3 The view update framework

In this section, we present a general framework for view updating based on the notion of determinacy under constraints introduced in Section 2. First, we revisit the definitions of translation and translatability that are given in [1]. Next, we show that a view induced by constraints is invertible exactly when the view symbols determine the database symbols under the given constraints. We then provide a general criterion to establish whether a view update can be successfully propagated to the underlying database.

We start by introducing a simple example that will be used throughout this section in order to illustrate the definitions and the results we present. We use the standard concise syntax for database dependencies and their FOL representation interchangeably (see [5] for the correspondence between the two formalisms).

*Example 1.* Consider the database schema $\mathcal{R} = \{R\}$, with $R$ over attributes $E$ for Employee, $D$ for Department and $M$ for Manager, in this order,[2] so that the first position of $R$ corresponds to $E$, the second to $D$ and the third to $M$, and let $\Sigma_\mathcal{R}$ consist of the fd $D \to M$. Let $\mathcal{V} = \{V_1, V_2\}$ with $V_1$ and $V_2$ defined by projections on

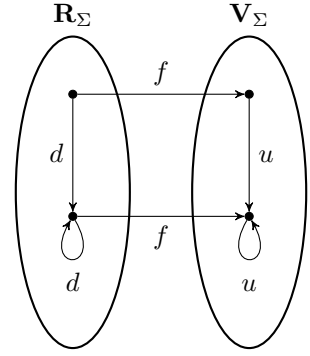

**Figure 1:** The database update $d$ is a translation of the view update $u$.

$ED$ and $DM$, respectively. That is, $\Sigma_{\mathcal{R}\mathcal{V}}$ consists of the following formulae:

$$\forall x, y\ V_1(x, y) \leftrightarrow \exists z\ R(x, y, z)\ ,\qquad(1\text{a})$$
$$\forall x, y\ V_2(x, y) \leftrightarrow \exists z\ R(z, x, y)\ .\qquad(1\text{b})$$

Then, let the set of global constraints be $\Sigma = \Sigma_\mathcal{R} \cup \Sigma_{\mathcal{R}\mathcal{V}}$.

A *database update* (respectively, *view update*) is a function $d\colon \mathbf{R} \to \mathbf{R}$ (resp., $u\colon \mathbf{V} \to \mathbf{V}$) associating each database state (resp., view state) with another, possibly the same. In what follows, we assume the presence of an underlying view $f$. Given a view update, we want to find a suitable database update that modifies the base relations so as to reflect *exactly* the changes of the view relations. In other words, the view update needs to be *translated* into a database update that brings the database to a new state, from which it is possible to reach the updated view state by means of the view mapping. In addition, unnecessary and unjustified changes in the database are to be avoided, in the sense that if the view update does not modify the view state, then the database update must not modify the corresponding database state either. Such requirements are formalised below (cf. Definition 3.1 in [1]) and exemplified in Figure 1.

*Definition 3 (Translation).* A database update $d$ is a *translation* of a view update $u$ if (1) $uf = fd$, and (2) for every $I_\mathcal{R} \in \mathbf{R}_\Sigma$, $d(I_\mathcal{R}) = I_\mathcal{R}$ whenever $uf(I_\mathcal{R}) = f(I_\mathcal{R})$.

A necessary condition for a view update to have a translation is that the update leads to view states that are reachable from some database state by means of $f$.

*Definition 4 (Translatability).* A view update $u$ has a translation, that is, is *translatable*, if for each $I_\mathcal{R} \in \mathbf{R}_\Sigma$ there is $I'_\mathcal{R} \in \mathbf{R}_\Sigma$ such that $f(I'_\mathcal{R}) = uf(I_\mathcal{R})$.

Translatability of view updates ensures that there exists a translation, but it does not rule out the possibility that there might be more than one. To avoid ambiguity, we are only interested in view updates for which there is one and only one translation, and we call such view updates *uniquely translatable*. We restrict our attention to injective views, for which it turns out that a view update is translatable iff it is uniquely translatable.

---

[2]W.l.o.g. we assume sets of attributes to be totally ordered, in order to easily switch between the *named* (i.e., attribute-based) and *unnamed* (i.e., position-based) perspectives.

THEOREM 2. *Assume $f$ is injective. Then, every translatable view update is uniquely translatable.*

The following theorem provides a characterisation of the unique database update into which a translatable view update is translated when the view is invertible (hence injective).

THEOREM 3. *Let $f$ be invertible, let $u$ be a translatable view update and let $d$ be a database update. Then, $d$ is a translation of $u$ if and only if $d = f^{-1}uf$.*

We now show the first main result of this section, namely that views induced by constraints can be inverted exactly when the database symbols are determined by the view symbols under the given constraints. In this situation, the constraints induce two mappings, one from the database states to the view states and one in the opposite direction, which are indeed one the inverse of the other.

THEOREM 4. *$f$ is invertible if and only if $\mathcal{V} \twoheadrightarrow_\Sigma^h \mathcal{R}$, and in such a case $h = f^{-1}$.*

When $\mathcal{V}$ determines $\mathcal{R}$ under $\Sigma$, the extension of the database symbols functionally depends on that of the view symbols, but nothing is known on *how* the former is to be computed from latter. Indeed, in order to constructively characterise the inverse of a view induced by $\Sigma$, we need to be able to explicitly express each database symbol $R \in \mathcal{R}$ in terms of the view symbols $\mathcal{V}$ by means of a formula $\psi \in \mathfrak{L}$, called an *exact rewriting of $R$ in terms of $\mathcal{V}$ under $\Sigma$*, such that $\mathsf{sig}(\psi) \subseteq \mathcal{V}$ and $\Sigma \models \forall \overline{x} \left( R(\overline{x}) \leftrightarrow \psi(\overline{x}) \right)$. As in this paper we are only concerned with exact rewritings, we sometimes simply call them "rewritings". We discuss different settings in which determinacy can be checked, and rewritings effectively found, in the upcoming Sections 4, 5 and 6.

Assuming such a constructive characterisation of $f^{-1}$ is obtained, whenever a view update leads to a view state $I_\mathcal{V}$ in $\mathbf{V}_\Sigma$, the changes can be propagated to the database state $f^{-1}(I_\mathcal{V})$ by computing the extension of each database symbol from its rewriting in terms of the view symbols. However, how do we know whether a view state belongs in fact to $\mathbf{V}_\Sigma$? Let $\widetilde{\Sigma}_\mathcal{V}$ be obtained from $\Sigma$ by replacing every occurrence of each $R \in \mathcal{R}$ with its rewriting in terms of $\mathcal{V}$. The resulting set of constraints, which we call the *$\mathcal{V}$-embedding* of $\Sigma$, mentions only view symbols and, as it turns out, is satisfied exactly by all and only the view states in $\mathbf{V}_\Sigma$. We can then prove the other main result of this section, namely that checking whether the changes introduced by a view update can be (uniquely) translated amounts to checking whether the view state resulting from the view update satisfies such view constraints $\widetilde{\Sigma}_\mathcal{V}$.

THEOREM 5. *Let $\mathcal{V} \twoheadrightarrow_\Sigma \mathcal{R}$ and let $u$ be a view update. Then, $u$ is translatable iff $u(I_\mathcal{V}) \models \widetilde{\Sigma}_\mathcal{V}$ for every $I_\mathcal{V} \in \mathbf{V}_\Sigma$.*

In other words, a view update is translatable if and only if it brings each view state that is legal w.r.t. $\widetilde{\Sigma}_\mathcal{V}$ to another such view state.

A view update can be expressed by means of an appropriate set $\Xi$ of constraints over the signature $\mathcal{V} \cup \mathsf{ren}(\mathcal{V})$, where $\mathsf{ren}$ is a renaming over $\mathcal{V}$ and $\mathcal{V} \twoheadrightarrow_\Xi \mathsf{ren}(\mathcal{V})$. Intuitively, the symbols in $\mathsf{ren}(\mathcal{V})$ represent the view schema *after* the update and, under the constraints defining the view update, they are determined by the symbols in the view schema $\mathcal{V}$ *before* the update. Refer to [13] for further technical details about the above representation of view updates by means of constraints.

For a view symbol $V \in \mathcal{V}$, let $V' = \mathsf{ren}(V)$. The insertion into $V$ and the deletion from $V$ of a tuple $\overline{x}$ are represented by the following open formulae:

$\mathsf{insert}_V(\overline{x}) \equiv \forall \overline{y} \left[ V'(\overline{y}) \leftrightarrow \left( V(y) \vee \overline{y} = \overline{x} \right) \right]$ ;

$\mathsf{delete}_V(\overline{x}) \equiv \forall \overline{y} \left[ \left( V'(\overline{y}) \rightarrow V(\overline{y}) \right) \wedge \right.$
$\left. \left( V(\overline{y}) \rightarrow \left[ V'(\overline{y}) \vee \overline{y} = \overline{x} \right] \right) \wedge \neg V'(\overline{x}) \right]$ .

Any update that does not modify $V$ can be represented by the closed formula $\mathsf{noop}_V \equiv \forall \overline{x} \left( V'(\overline{x}) \leftrightarrow V(\overline{x}) \right)$, while the replacement of a tuple $\overline{x}$ with a tuple $\overline{y}$ is expressed by the following open formula:

$\mathsf{repl}_V(\overline{x}, \overline{y}) \equiv \left[ \left( \neg V(\overline{x}) \vee \overline{x} = \overline{y} \right) \rightarrow \mathsf{noop}_V \right] \wedge$
$\left[ \left( V(\overline{x}) \wedge \overline{x} \neq \overline{y} \right) \rightarrow \mathsf{delete}_V(\overline{x}) \wedge \mathsf{insert}_V(\overline{y}) \right]$ .

Transactional updates, in the sense of sequences of updates applied one after the other, are also expressible in our formalism. For example, the update $\mathsf{insert}_V(\overline{a}) \wedge \mathsf{delete}_{V'}(\overline{b})$ represents the insertion of $\overline{a}$ into $V$ followed by the deletion of $\overline{b}$. Indeed, $\mathsf{delete}_{V'}(\overline{b})$ is applied on $V'$, which is the result of applying $\mathsf{insert}_V(\overline{a})$ on $V$.

From Theorem 5, we get the following characterisation of the translatability of view updates in terms of logical implication.

THEOREM 6. *Let $f$ be invertible and $u$ be a view update expressed by $\Xi$. Then, $u$ is translatable iff $\widetilde{\Sigma}_\mathcal{V} \cup \Xi \models \mathsf{ren}(\widetilde{\Sigma}_\mathcal{V})$.*

*Example 2 (cont'd from Example 1).* It can be checked that $R(x, y, z)$ has the rewriting $V_1(x, y) \wedge V_2(y, z)$. Therefore, the $\mathcal{V}$-embedding of $\Sigma$ consists of the inclusion dependencies $V_1[D] \subseteq V_2[D]$ and $V_2[D] \subseteq V_1[D]$, and of the functional dependency $V_2 \colon D \rightarrow M$. Then, the conditional view update that inserts $\langle e, d \rangle$ into $V_1$ only if there is already another tuple with the same value $d$ for attribute $D$, and does nothing otherwise, is translatable.

Note that a view update which is not translatable in general might still be translated when applied on a given view state. Checking whether an update is *translatable w.r.t. a view state $I_\mathcal{V}$* satisfying $\widetilde{\Sigma}_\mathcal{V}$ (i.e., whether $u(I_\mathcal{V}) \models \widetilde{\Sigma}_\mathcal{V}$) can be done in PTIME in the size of $u(I_\mathcal{V})$, which is the data complexity of testing whether a finite relational structure is a model of a FOL theory.

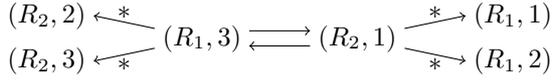

**Figure 2:** Dependency graph for $\Sigma_{\mathcal{R}}$ of Example 4.

*Example 3 (continued from Example 2).* Suppose the extension of $R$ is $\{\langle e, d, m\rangle\}$. In turn, the extensions of $V_1$ and $V_2$ are $\{\langle e, d\rangle\}$ and $\{\langle d, m\rangle\}$, respectively. Then, the insertion of $\langle e, d'\rangle$ into $V_1$ is rejected as it violates the inclusion dependency between $V_1$ and $V_2$ on $D$, while the simultaneous insertion of $\langle e, d'\rangle$ into $V_1$ and $\langle d', m\rangle$ into $V_2$ (achieved through a transactional update) is accepted and translated as the insertion of $\langle e, d', m\rangle$ into $R$.

## 4 Conjunctive Views

The first setting we investigate is one where the view symbols are defined by arbitrary conjunctive queries (CQs) over a multi-relational database schema and the database constraints are stratified embedded dependencies. Embedded dependencies are expressive enough to capture virtually all other classes of dependencies studied in the literature [5].

*Example 4.* Let $\mathcal{R} = \{R_1, R_2\}$ and $\Sigma_{\mathcal{R}}$ consist of the following embedded dependencies (in this case, inclusion and functional dependencies):[3]

$$R_1(x, y, z) \to \exists x', y'\ R_2(z, x', y')\ , \quad (2a)$$
$$R_2(x, y, z) \to \exists y', z'\ R_1(y', z', x)\ , \quad (2b)$$
$$R_1(x, y, z) \land R_1(x', y, z') \to z = z'\ , \quad (2c)$$
$$R_2(x, y, z) \land R_2(x, y', z') \to y = y'\ . \quad (2d)$$

Let $\mathcal{V} = \{V_1, V_2, V_3\}$ and $\Sigma_{\mathcal{R}\mathcal{V}}$ consist of the following view definitions:

$$V_1(x, y) \leftrightarrow \exists z\ \ R_1(x, y, z)\ , \quad (3a)$$
$$V_2(x, y, z) \leftrightarrow \exists v, w\ R_1(v, x, y) \land R_2(y, z, w)\ , \quad (3b)$$
$$V_3(x, y) \leftrightarrow \exists z\ \ R_2(x, z, y)\ . \quad (3c)$$

We consider embedded dependencies that are required to be *stratified* (refer to [6] for the formal definition). This notion is based on the *chase graph*: the chase graph of a set of embedded dependencies $\Sigma$ has the constraints in $\Sigma$ as nodes and, for $\alpha, \beta \in \Sigma$, there is an edge from $\alpha$ to $\beta$ iff, intuitively, firing $\alpha$ may cause $\beta$ to fire. Then, a set of embedded dependencies $\Sigma$ is *stratified* if the set of constraints in every cycle of its chase graph is *weakly acyclic* [9, 8]. Note that every weakly acyclic set of dependencies is also stratified.

*Example 5 (cont'd from Example 4).* The *dependency graph* [9] for $\Sigma_{\mathcal{R}}$, shown in Figure 2, does not contain any cycle going through a special edge, therefore $\Sigma_{\mathcal{R}}$ is weakly acyclic and, in turn, stratified.

---
[3]Universal quantifiers are omitted.

The main result of this section establishes that, when the embedded dependencies over the database schema are stratified, the view is invertible iff each database symbol has an exact rewriting in terms of the view symbols under the global constraints given by a conjunctive query.

THEOREM 7. *Let $\Sigma_{\mathcal{R}}$ be a set of stratified embedded dependencies, let $\Sigma_{\mathcal{R}\mathcal{V}}$ be such that the definition of each $V \in \mathcal{V}$ is given by a CQ over $\mathcal{R}$, and let $\Sigma = \Sigma_{\mathcal{R}} \cup \Sigma_{\mathcal{R}\mathcal{V}}$. Then, $\mathcal{V} \twoheadrightarrow_\Sigma \mathcal{R}$ iff each $R \in \mathcal{R}$ has an exact CQ rewriting in terms of $\mathcal{V}$ under $\Sigma$.*

The *Chase and Backchase* (C&B) [7] is an algorithm that enumerates the exact (conjunctive) rewritings of a CQ under constraints. More precisely, given two schemas $\mathcal{S}$ and $\mathcal{T}$, a set of embedded dependencies $\Gamma$ over $\mathcal{S} \cup \mathcal{T}$, and an input CQ $q$ over $\mathcal{S}$, the C&B outputs the CQs over $\mathcal{T}$ which are equivalent to $q$ under $\Gamma$. The C&B is sound and complete, in the sense that it returns all and only the CQs into which the input CQ can be rewritten under the given constraints, whenever the chase is guaranteed to terminate, which is the case, e.g., for stratified sets of dependencies. Obviously, the fact that the output of the C&B is empty for $q$ does not mean that $q$ has no rewriting in terms of $\mathcal{T}$ under $\Gamma$, but simply that its rewriting, if any, is not a CQ.

We can use the C&B in order to obtain the rewritings we are interested in. For each $R \in \mathcal{R}$, consider the atomic query $q(\overline{x})$ :- $R(\overline{x})$ and proceed in the following two phases:

**Chase.** Chase $q$ with $\Sigma$ until no further chase step applies. The resulting query is the so-called *universal plan* $U$.

**Backchase.** Every subquery of $U$ over $\mathcal{V}$ (i.e., a set of $\mathcal{V}$-atoms from $U$ mentioning all of $q$'s free variables) is a candidate for being a rewriting of $q$. Chase each candidate $q'$ with $\Sigma$ step-by-step until no further chase step applies, and at each new step in the chase sequence check whether a containment mapping from the original query $q$ can be found. If that is the case, then $q'$ is a rewriting of $q$.

The above is described in more detail in Algorithms 1 and 2, which we show to be sound and complete below and which we then illustrate on our running example.

THEOREM 8. *Let $\Sigma$ be as in Theorem 7. Then, the procedure REWRITE of Algorithm 1 is sound and complete for finding the rewriting of each database symbol in terms of the view symbols under $\Sigma$, and DETERMINES of Algorithm 2 is a sound and complete procedure for deciding whether $\mathcal{V} \twoheadrightarrow_\Sigma \mathcal{R}$.*

*Example 6 (continued from Example 5).* Chasing the query $q(x, y, z)$ :- $R_1(x, y, z)$ with $\Sigma$ yields the universal plan:

$$U(x, y, z) \text{ :- } R_1(x, y, z), R_2(z, x', y'),$$
$$V_1(x, y), V_3(z, y'), V_2(y, z, x')\ . \quad (4)$$

**Algorithm 1**

**Input:**
 a) an atomic query over $\mathcal{R}$,
 b) a view schema $\mathcal{V}$,
 c) a set of global constraints $\Sigma$ over $\mathcal{R} \cup \mathcal{V}$.

**Output:** an exact CQ rewriting of $q$ in terms of $\mathcal{V}$ under $\Sigma$, if any, or $\bot$ otherwise.

1: **function** REWRITE($q, \mathcal{V}, \Sigma$)
2:    $U = \mathsf{chase}(q, \Sigma)$
3:    **for** each subquery $q'$ of $U$ over $\mathcal{V}$ **do**
4:       candidate $= q'$
5:       **repeat**
6:          $q' = \mathsf{chase\text{-}step}(q', \Sigma)$
7:          **if** $\exists$ containment mapping from $q$ to $q'$ **then**
8:             **return** candidate
9:          **end if**
10:      **until** no further chase step applies
11:   **end for**
12:   **return** $\bot$
13: **end function**

A candidate for being a rewriting of $R(x,y,z)$ in terms of $\mathcal{V}$ is the subquery $q'(x,y,z) \text{ :- } V_1(x,y), V_2(y,z,x')$, that chased with the left-to-right tgd's from (3a) and (3b) gives:

$$q''(x,y,z) \text{ :- } V_1(x,y),\ V_2(y,z,x'), \\ R_1(x,y,z'),\ R_1(v,y,z),\ R_2(z,x',w)\ . \quad (5)$$

A chase step with (2c) yields $z' = z$, and we can thus find a containment mapping (the identity) from the original query $q$ to (5). Therefore, the rewriting of $R_1(x,y,z)$ is $\exists w\ V_1(x,y) \wedge V_2(y,z,w)$. Similarly, we also have that $R_2(x,y,z)$ can be rewritten in terms of $\mathcal{V}$ as $\exists v\ V_2(v,x,y) \wedge V_3(x,z)$.

**Algorithm 2**

**Input:**
 a) a view schema $\mathcal{V}$,
 b) a database schema $\mathcal{R}$,
 c) a set of global constraints $\Sigma$ over $\mathcal{R} \cup \mathcal{V}$.

**Output:** True if $\mathcal{V} \twoheadrightarrow_\Sigma \mathcal{R}$, False otherwise.

1: **procedure** DETERMINES($\mathcal{V}, \mathcal{R}, \Sigma$)
2:   **for** each $R \in \mathcal{R}$ **do**
3:      $q$ is the atomic query $R(\overline{x})$
4:      **if** REWRITE($q, \mathcal{V}, \Sigma$) $= \bot$ **then**
5:         **return** False
6:      **end if**
7:   **end for**
8:   **return** True
9: **end procedure**

The result presented here settles a long-standing open issue pointed out in [3], namely how to solve the view update problem in a multi-relational database with views that are projections of joins of relations, and extends the setting of [12], consisting of *only one* database symbol, view symbols defined by acyclic projections and database constraints given by full dependencies, which is a special case where the rewriting is known to be the join (rather than a general CQ).

## 5 Horizontal Decompositions

We now turn our attention to a setting where the view symbols are defined by selections over a single-relation database with a specific kind of domain constraints. In particular, we consider a database schema where some attributes are *interpreted*, that is, they take their values in specific domains, such as the integers or the reals, on which a set of predicates and functions are defined, according to a constraint language $\mathfrak{C}$. We consider database constraints in $\Sigma_\mathcal{R}$ (with $\mathcal{R} = \{R\}$) of the form:

$$\forall \overline{x}, \overline{y}\ .\ R(\overline{x}, \overline{y}) \wedge \overline{x}' = \overline{a} \wedge \overline{x}'' \neq \overline{b} \to \delta(\overline{y})\ , \quad (6)$$

with $\delta(\overline{y}) \in \mathfrak{C}$. Each view symbol $V \in \mathcal{V}$ is defined in $\Sigma_{\mathcal{R}\mathcal{V}}$ by:

$$\forall \overline{x}, \overline{y}\ .\ V(\overline{x}, \overline{y}) \leftrightarrow \bigl(R(\overline{x}, \overline{y}) \wedge \overline{x}' = \overline{a} \wedge \overline{x}'' \neq \overline{b} \wedge \sigma(\overline{y})\bigr)\ , \quad (7)$$

with $\sigma(\overline{y}) \in \mathfrak{C}$.

W.l.o.g. we assume that the first $k$ positions of $R$ correspond to non-interpreted attributes (denoted by the $\overline{x}$ variables), while the others correspond to interpreted ones (denoted by the $\overline{y}$ variables); $\overline{x}'$ and $\overline{x}''$ denote disjoint subsets of the variables in $\overline{x}$. A variable appearing in the $i$-th position of $R$ is named $x_i$ if $i \leq k$ (i.e., that position corresponds to a non-interpreted attribute) and $y_{i-k}$ otherwise. Clearly, this is w.l.o.g. as it can be easily achieved by renaming.

*Example 7.* Let $R$ be on attributes Name, Department, Position, Salary and Bonus, in this order, where the last two are interpreted over the integers. Let $\Sigma_\mathcal{R}$ consist of the following database constraints:

$$R(\overline{x}, \overline{y}) \wedge x_2 = \text{``ICT''} \quad \to y_1 + y_2 \leq 5\ , \quad (8a)$$
$$R(\overline{x}, \overline{y}) \wedge x_3 = \text{``Manager''} \to y_2 \qquad \geq 2\ , \quad (8b)$$
$$R(\overline{x}, \overline{y}) \qquad \qquad \qquad \to y_1 - y_2 \geq 0\ . \quad (8c)$$

Intuitively, the above constraints state that (8a) employees working in the ICT department get a total income (consisting of salary plus bonus) of at most 5, say, thousands of euros per month; that (8b) employees who work as managers get a bonus of at least 2, and that (8c) employees get a bonus not greater than their salary.

Let $\mathcal{V} = \{V_1, V_2, V_3\}$ and let $\Sigma_{\mathcal{R}\mathcal{V}}$ define the view symbols by means of the following formulae:

$$V_1: R(\overline{x}, \overline{y}) \wedge x_2 \neq \text{``ICT''} \wedge x_3 = \text{``Manager''}\ , \quad (9a)$$
$$V_2: R(\overline{x}, \overline{y}) \wedge y_2 < 4\ , \quad (9b)$$
$$V_3: R(\overline{x}, \overline{y}) \wedge x_3 \neq \text{``Manager''}\ . \quad (9c)$$

Intuitively, $V_1$, $V_2$ and $V_3$ select all employees who (9a) work as managers in departments other than ICT, who (9b) get a bonus strictly smaller than 4, and who (9c) do not work as managers, respectively.

Determinacy under constraints (therefore, invertibility of views) in the setting studied in this section is strongly connected with the notion of *horizontal decomposition* [4], that we formally define below.

*Definition 5 (Horizontal decomposition).* Let $\mathcal{R} = \{R\}$ be a database schema, $\mathcal{V} = \{V_1, \ldots, V_n\}$ be a view schema such that $|V_i| = |R|$ for each $V_i \in \mathcal{V}$, and let $\Sigma$ be a set of global constraints over $\mathcal{R} \cup \mathcal{V}$. We say that $V_1, \ldots, V_n$ form a *horizontal decomposition* of $R$ under $\Sigma$ if $\Sigma \models \forall \overline{x}\ V_i(\overline{x}) \to R(\overline{x})$ for every $V_i \in \mathcal{V}$. In addition, a horizontal decomposition is said to be *lossless* if $\Sigma \models \forall \overline{x}\ R(\overline{x}) \leftrightarrow V_1(\overline{x}) \vee \cdots \vee V_n(\overline{x})$.

The first result of this section establishes that, when the database constraints are of the form (6) and the view symbols are defined as in (7), the view is invertible iff the horizontal decomposition formed by the view symbols is lossless.

THEOREM 9. *Let $\Sigma = \Sigma_\mathcal{R} \cup \Sigma_{\mathcal{RV}}$, where $\Sigma_\mathcal{R}$ consists of constraints of the form (6) and the view definitions in $\Sigma_{\mathcal{RV}}$ are as in (7). Then, $\mathcal{V} \twoheadrightarrow_\Sigma \mathcal{R}$ iff the view symbols in $\mathcal{V}$ form a lossless horizontal decomposition of $R$ under $\Sigma$, that is, iff $\Sigma \models \forall \overline{x}\ R(\overline{x}) \leftrightarrow V_1(\overline{x}) \vee \cdots \vee V_n(\overline{x})$.*

In other words, the above theorem says that the view symbols in $\mathcal{V}$ determine the database symbol $R$ under $\Sigma$ iff $R$ has an exact rewriting in terms of $\mathcal{V}$ under $\Sigma$ which is given by the union of the view symbols. However, the problem of effectively checking whether $R$ has indeed such a rewriting is not addressed by Theorem 9. We do so next.

Let the propositional variable $p_i^a$ indicate that the value in the $i$-th position is $a$. With each $\psi \in \Sigma$, we associate a propositional formula $P(\psi)$ which we call the *propositional representation* of $\psi$. The propositional representation of $\psi \in \Sigma_\mathcal{R}$ is

$$\Big[ \bigwedge_{\substack{x_i \in \mathsf{var}(\overline{x}') \\ x_i = \overline{x}'[j]}} p_i^{\overline{a}[j]} \Big] \wedge \Big[ \bigwedge_{\substack{x_i \in \mathsf{var}(\overline{x}'') \\ x_i = \overline{x}''[j]}} \neg p_i^{\overline{b}[j]} \Big] \to v \ , \qquad (10)$$

in which $v$ is a fresh propositional variable associated with the constraint $\delta(\overline{y})$ appearing in $\psi$. Similarly, the propositional representation of each $\psi \in \Sigma_{\mathcal{RV}}$ is

$$\Big[ \bigwedge_{\substack{x_i \in \mathsf{var}(\overline{x}') \\ x_i = \overline{x}'[j]}} p_i^{\overline{a}[j]} \Big] \wedge \Big[ \bigwedge_{\substack{x_i \in \mathsf{var}(\overline{x}'') \\ x_i = \overline{x}''[j]}} \neg p_i^{\overline{b}[j]} \Big] \wedge v \ , \qquad (11)$$

in which $v$ is a fresh propositional variable associated with the constraint $\sigma(\overline{y})$, if any, appearing in $\psi$. Let

$$\Gamma_\delta = \{\ P(\psi) \mid \psi \in \Sigma_\mathcal{R}\ \} \ , \qquad (12)$$
$$\Gamma_\sigma = \{\neg P(\psi) \mid \psi \in \Sigma_{\mathcal{RV}}\} \ , \qquad (13)$$

where $P(\psi)$ is the propositional representation of $\psi$ as above, and let

$$\Gamma_{\mathrm{ax}} = \big\{ \neg(p_i^a \wedge p_i^b) \mid a \neq b,\ p_i^a, p_i^b \in \mathsf{pvar}(\Gamma) \big\} \ , \qquad (14)$$

intuitively stating that distinct values are not allowed in the same position.

We call the set $\Gamma = \Gamma_\delta \cup \Gamma_\sigma$ the *propositional theory associated with* $\Sigma$, and $\Gamma_{\mathrm{ax}}$ the corresponding *distinctness axioms* for it. For $\Gamma' \subseteq \Gamma$, let $\mathsf{cvar}(\Gamma')$ be the set of propositional variables occurring in $\Gamma'$, and let $\mathsf{cvar}(\Gamma')$ be the set of propositional variables in $\mathsf{var}(\Gamma')$ having an associated $\mathfrak{C}$-constraint, denoted by $\mathsf{constr}(v)$ for each $v \in \mathsf{cvar}(\Gamma')$. We use $\mathsf{pvar}(\Gamma')$ as short for $\mathsf{var}(\Gamma') \setminus \mathsf{cvar}(\Gamma)$.

*Example 8 (cont'd from Example 7).* For the sake of simplicity, let $a =$ "ICT" and $b =$ "Manager". The sets $\Gamma_\delta$ and $\Gamma_\sigma$ corresponding to (8a)–(8c) and (9a)–(9c), respectively, are

$$\Gamma_\delta = \{\ p_2^a \to v_1,\ p_3^b \to v_2,\ \top \to v_3\ \} \ , \qquad (15)$$
$$\Gamma_\sigma = \{\ p_2^a \vee \neg p_3^b,\ \neg v_4,\ p_3^b\ \} \ , \qquad (16)$$

and $\Gamma_{\mathrm{ax}} = \varnothing$. The set of propositional variables occurring in $\Gamma = \Gamma_\delta \cup \Gamma_\sigma$ is $\mathsf{var}(\Gamma) = \{p_2^a, p_3^b, v_1, v_2, v_3, v_4\}$ and the set $\mathsf{cvar}(\Gamma) \subseteq \mathsf{var}(\Gamma)$ of those associated with a $\mathfrak{C}$-constraint is $\{v_1, v_2, v_3, v_4\}$, where the association $\mathsf{constr}$ is given by:

$$\{\quad v_1 \mapsto y_1 + y_2 \leq 5, \qquad v_2 \mapsto y_2 \geq 2,$$
$$\quad v_3 \mapsto y_1 - y_2 \geq 0, \qquad v_4 \mapsto y_2 < 4 \quad \} \ .$$

Given a valuation $\alpha$ on $\mathsf{var}(\Gamma)$, let $\Gamma^\alpha = \Gamma_\delta^\alpha \cup \Gamma_\sigma^\alpha$ be the set of $\mathfrak{C}$-constraints where

$$\Gamma_\delta^\alpha = \{\mathsf{constr}(v) \mid v \in \mathsf{cvar}(P),$$
$$\qquad P = (L \to v) \in \Gamma_\delta,\ \alpha(L) = \mathtt{T}\} \ , \quad (17)$$

that is, the $\mathfrak{C}$-constraints associated with propositional variables occurring in some formula of $\Gamma_\delta$ whose l.h.s. holds true under $\alpha$, and

$$\Gamma_\sigma^\alpha = \{\quad \mathsf{constr}(v) \mid v \in \mathsf{cvar}(\Gamma_\sigma),\ \alpha(v) = \mathtt{T}\} \cup$$
$$\qquad \{\neg\,\mathsf{constr}(v) \mid v \in \mathsf{cvar}(\Gamma_\sigma),\ \alpha(v) = \mathtt{F}\} \ , \quad (18)$$

that is, the $\mathfrak{C}$-constraints associated with propositional variables in $\Gamma_\sigma$, taken positively or negatively depending on the truth value assigned to them by $\alpha$.

We are now ready to state the main result of this section, that characterises determinacy under constraints (i.e., equivalently, losslessness of horizontal decompositions) in terms of satisfiability of the propositional theory associated with the global constraints and unsatisfiability in $\mathfrak{C}$.

THEOREM 10. *Let $\Sigma$ be as in Theorem 9, let $\Gamma$ be the propositional theory associated with $\Sigma$ and let $\Gamma_{\mathrm{ax}}$ be the distinctness axioms for $\Gamma$. Then, $\Sigma \models \forall \overline{x}\ R(\overline{x}) \leftrightarrow V_1(\overline{x}) \vee \cdots \vee V_n(\overline{x})$ iff $\Gamma^\alpha$ is unsatisfiable for every valuation $\alpha$ satisfying $\Gamma \cup \Gamma_{ax}$.*

When the constraint language $\mathfrak{C}$ is closed under negation and the satisfiability of $\mathfrak{C}$-formulae is decidable, Theorem 10 directly yields an algorithm for deciding whether $\mathcal{V} \twoheadrightarrow_\Sigma \mathcal{R}$. This is the case, e.g., for boolean combinations of *Unit Two Variable Per Inequality* (UTVPI) constraints [22], where a UTVPI has the form $ax + by \leq d$, with $x$ and $y$ integer variables, $a, b \in \{-1, 0, 1\}$ and $d \in \mathbb{Z}$. UTVPI constraints are a fragment of linear arithmetic constraints. We conclude the section by illustrating the algorithm for checking determinacy on our running example, which can indeed be expressed using UTVPI constraints.

*Example 9 (continued from Example 8).* The only valuation satisfying $\Gamma$ is

$$\alpha = \{\quad p_2^a \mapsto \mathtt{T}, \quad p_3^b \mapsto \mathtt{T}, \quad v_1 \mapsto \mathtt{T},$$
$$v_2 \mapsto \mathtt{T}, \quad v_3 \mapsto \mathtt{T}, \quad v_4 \mapsto \mathtt{F} \quad \} \ .$$

Hence, we have:

$$\Gamma_\delta^\alpha = \{y_1 + y_2 \leq 5,\ y_2 \geq 2,\ y_1 - y_2 \geq 0\} \ , \quad (19)$$
$$\Gamma_\sigma^\alpha = \{y_2 \geq 4\} \ . \quad (20)$$

Note that the constraint in (20) is $\neg\,\mathsf{constr}(v_4)$, that is, the negation of $y_2 < 4$. The set $\Gamma^\alpha = \Gamma_\delta^\alpha \cup \Gamma_\sigma^\alpha$ can be shown to be unsatisfiable as follows: from $y_1 + y_2 \leq 5$ and $y_2 \geq 4$ we get $y_1 \geq 4$, or equivalently $-y_1 \leq -4$, which together with $y_1 + y_2 \leq 5$ yields $y_2 \leq 1$, in conflict with $y_2 \geq 2$. Therefore, $\mathcal{V} \twoheadrightarrow_\Sigma \mathcal{R}$.

## 6 Ontological constraints

In this section, we investigate another multi-relational database setting, where views are defined in an expressive fragment of relational algebra (RA), called *Type Relational Algebra*, and database constraints are inclusions of expressions in such an algebra.

Expressions in the Type RA are relational algebra expressions with a restricted use of projections and cross products, as specified below:

- arbitrary boolean RA expressions (union, intersection, difference);
- unary projections of the form $\pi_i s$ where $s$ is a Type RA expression of arity $n \geq 2$ and $i \leq n$;
- and so called *typed selections* of arity $n$ of the form

$$\underbrace{\mathbf{dom} \times \cdots \mathbf{dom}}_{i-1} \times\ t\ \times \underbrace{\mathbf{dom} \cdots \times \mathbf{dom}}_{n-i} \ ,$$

written $\sigma_{i/n} t$, where $t$ is a Type RA expression of arity 1 and $i \leq n$.

*Example 10.* Let $\mathcal{R} = \{\mathsf{Person}, \mathsf{Name}, \mathsf{DOB}, \mathsf{Citizenship}, \mathsf{EU}, \mathsf{non\text{-}EU}\}$ and let $\Sigma_\mathcal{R}$ consist of the following Type RA inclusions:

$$\mathsf{Person} \subseteq \sigma_{1/3}\mathsf{Name} \cap \sigma_{2/3}\mathsf{DOB} \cap \sigma_{3/3}\mathsf{Citizenship}$$
$$\mathsf{Citizenship} = \mathsf{EU} \cup \mathsf{non\text{-}EU}$$
$$\varnothing \supseteq \mathsf{EU} \setminus \mathsf{non\text{-}EU}$$
$$\varnothing \supseteq \mathsf{non\text{-}EU} \setminus \mathsf{EU}$$

Here, a Person is a ternary relation with the first argument typed to be a Name, the second one typed to be a DOB, and the third typed to be a Citizenship; moreover, a Citizenship is a partition of the EU and non-EU countries.

Let $\mathcal{V} = \{\mathsf{European}, \mathsf{Extra\text{-}European}\}$ and let $\Sigma_{\mathcal{RV}}$ consist of the following Type RA view definitions:

$$\mathsf{European} = \mathsf{Person} \cap \sigma_{3/3}\mathsf{EU}$$
$$\mathsf{Extra\text{-}European} = \mathsf{Person} \setminus \sigma_{3/3}\mathsf{EU}$$

The first view defines all the people having a European citizenship, while the second one defines all the people who do not have a European citizenship.

THEOREM 11. *Let $\Sigma_\mathcal{R}$ and $\Sigma_{\mathcal{RV}}$ consist of constraints and view definitions in Type RA as specified above, and let $\Sigma = \Sigma_\mathcal{R} \cup \Sigma_{\mathcal{RV}}$. Then, $\mathcal{V} \twoheadrightarrow_\Sigma \mathcal{R}$ if and only if for each $R \in \mathcal{R}$ the following decidable entailment problem holds:*

$$\Sigma \cup \mathsf{ren}(\Sigma) \models_{unr} \forall \overline{x}\ R(\overline{x}) \leftrightarrow \big(\mathsf{ren}(R)\big)(\overline{x}) \ , \quad (21)$$

*where ren is a renaming of $\mathcal{R}$, and $\models_{unr}$ denotes entailment under unrestricted (i.e., possibly infinite) instances.*

We are allowed to use unrestricted entailment here since (21) is finitely controllable. As a matter of fact, (21) can be reduced to satisfiability in a fragment of the $\mathcal{DLR}$ description logic [2], which is decidable and is in EXPTIME. This fragment is obtained by dropping cardinality constraints and limiting concept negation to concept difference in $\mathcal{DLR}$; by using arguments similar to [21], it can be shown that such a fragment has the finite model property. The reduction is the obvious one-to-one transformation from Type RA unary expressions to $\mathcal{DLR}$ concepts and from Type RA $n$-ary expressions ($n \geq 2$) to $\mathcal{DLR}$ $n$-ary relations; database constraints and view definitions are transformed to the corresponding $\mathcal{DLR}$ axioms. In order to better understand the expressivity of the Type RA as an ontology language, note that by dropping cardinality constraints and limiting concept negation to concept difference in $\mathcal{DLR}$, the reverse reduction also holds.

Theorem 11 is an application of the determinacy under constraints framework thoroughly analysed in [14]. Therefore, the exact rewriting of each database symbol in terms of the view symbols can be effectively obtained by means of interpolation-based techniques [14]. In our previous example, the exact rewriting of the database symbol Person is:

$$\mathsf{Person} = \mathsf{European} \cup \mathsf{Extra\text{-}European} \ ,$$

while the other database symbols are unaffected by changes to the views.

Of course, as specified by Theorem 5, one should check in addition that a specific update is translatable and that its (unique) translation satisfies the database constraints, e.g., in our case by not introducing unknown names or countries.

# 7 Related Work

As already mentioned, the understanding of the theoretical foundations of the view update problem is due to the seminal work of Bancilhon and Spyratos [1]. In addition to the notions of translation and translatability, that we revisited in Section 3, they also introduce the notions of *view complement* and *translation under constant complement*. Given a view that is lossy (i.e., does not preserve all of the informative content of the underlying database) a view complement is a second view that contains enough information to attain losslessness when combined with the original view. The constant complement principle prescribes that updates on the view must not influence, directly or indirectly, the content of the view complement. In this paper, we did not discuss the issues concerning these notions, but it suffices to say that the results of Section 3 about the invertibility of views and the translatability of view updates hold also in the presence of view complements [13].

Cosmadakis and Papadimitriou [3] study a restricted setting that consists of only two view symbols defined by projections over a uni-relational database schema. They show that, in the absence of view constraints, when the database constraints are functional and join dependencies, the view is invertible if and only if the database symbol can be rewritten as the join of the two view symbols. Their result has been generalised in [12] to an arbitrary number of view symbols defined by acyclic projections and database constraints consisting of full dependencies. Both [12] and [3] are special cases of the setting discussed in Section 4, where the rewriting is a priori known to be the join, and determinacy under constraints is finitely controllable (i.e., it holds or not independently on whether infinite instances are allowed). To the best of our knowledge, [3] is the only comprehensive work in which the abstract framework by Bancilhon and Spyratos is applied to a relational setting. Necessary and sufficient conditions are provided for the translatability of insertions, deletions and replacements w.r.t. an instance, when the database constraints are fd's. Our general criterion subsumes all these conditions, it applies for more expressive database constraints (namely, stratified embedded dependencies), can be checked locally on the view schema, and allows for more general classes of updates (e.g., conditional and transactional) rather than single insertions, deletions and replacements. In addition, the problem of checking translatability of view updates w.r.t. *every* view state is not addressed in [3], while our Theorem 6 provides a characterisation of translatability w.r.t. every view state in terms of logical implication. An example of a view update that is translatable in this sense is given in Example 2.

Nash et al. [20] investigate the related problem of deciding whether the answer to a query $q$ over a database schema $\mathcal{R}$ is determined by a set of view symbols $\mathcal{V}$ defined over $\mathcal{R}$. For view definitions and queries expressed in languages ranging from FOL to CQs, they systematically study whether determinacy is decidable and whether the query language is *complete* for rewritings, meaning that queries determined by views can be rewritten in terms of the view symbols by using the same language in which they are originally expressed. The notion of determinacy used in [20] differs from the one in this article in that we consider it w.r.t. a set of global constraints that, along with the view definitions, include also additional constraints over the database and the view schemas. Moreover, we are only interested in atomic queries, and rewritings that are general domain-independent function-free FOL formulae, rather than belonging to the same fragment in which queries are expressed. We remark that our result about determinacy of Section 4 is not contradicted by the fact that the class CQ of conjunctive queries is not complete for CQ-to-CQ rewritings [20]. Rather, in the absence of database constraints (i.e., $\Sigma_{\mathcal{R}} = \varnothing$), a corollary of Theorem 7 is that CQ is complete for AQ-to-CQ rewritings, where AQ denotes the class of atomic queries.

Fan et al. [11] study the connection between determinacy and invertibility of views w.r.t. a query. The setting is similar to that of [20], where the global constraints $\Sigma$ are given by exact view definitions only, and there are no constraints on either the database schema $\mathcal{R}$ nor the view schema $\mathcal{V}$. Recasting the definition in [11] to match our notation and terminology, the view $f$ induced by $\Sigma$ is said to be *invertible relative to* a query $q$ over $\mathcal{R}$ if there exists a query $q^{-1}$ such that $q(I_{\mathcal{R}}) = q^{-1}(f(I_{\mathcal{R}}))$ for every database instance $I_{\mathcal{R}}$. Intuitively, this means that $f$ (called a *transformation* in [11]) preserves, without any loss, the information selected by the query $q$. This is closely related to our notion of invertibility of a view: the view $f$ is invertible (in our sense) iff it is invertible relative to every atomic query over $\mathcal{R}$, that is, relative to $q(\overline{x}) = R(\overline{x})$, for each $R \in \mathcal{R}$. In other words, $f$ is lossless (i.e., invertible) when it preserves the information contained in each database relation, selected by the (atomic) query over the corresponding database symbol. In light of the strong connection between these two notions of invertibility, and between our notion of determinacy under constraints and determinacy in [20], as we show that invertibility of views coincides with determinacy under constraints, in [11] it is shown that invertibility relative to a query coincides with determinacy in the sense of [20].

The framework we presented in Section 3 is reminiscent of a Data Exchange (DE) [9] setting. In DE the connection between source and target schemas is commonly expressed by means of so-called *source-to-target* tgd's (s-t tgd's), that is, tgd's where the l.h.s. of the implication mentions only source symbols while the r.h.s. mentions only target symbols. A *schema mapping* in DE is a triple $\langle \mathcal{S}, \mathcal{T}, \Gamma \rangle$, where $\mathcal{S}$ is the source schema, $\mathcal{T}$ is the target schema and $\Gamma$ is a set of s-t tgd's over $\mathcal{S} \cup \mathcal{T}$. A schema mapping in this sense is not a view mapping in our sense, but it rather represents a class of them, namely that of all the views under $\Gamma$, according to Definition 1. Indeed, for a given source instance $I_{\mathcal{S}}$, in general there can be more than one target instance $I_\mathcal{T}$ such that their disjoint union $I_\mathcal{S} \uplus I_\mathcal{T}$ (see Section 2) is a model of $\Gamma$, therefore $\mathcal{S} \not\rightarrow_\Gamma \mathcal{T}$. On the contrary, by considering exact view definitions, we ensure a one-to-one correspondence between views and sets of global constraints inducing them (up to logical equivalence). Clearly, also the notion of inverse of a schema mapping in DE differs from the standard notion of inverse function in the mathematical sense (and ours). Indeed, no schema mapping in DE has a unique inverse, although under certain circumstances there is, among the others, only one inverse with a specific form [10]. The case in which there is a unique target instance for a given source instance under the schema mapping, namely when schema mappings can be transformed into proper view mappings via an abduction process, is analysed in [15].

# 8 Outlook

We conclude by pointing out and discussing several possible directions in the line of research of this paper that would be interesting to pursue and investigate further.

1. Consider views defined by queries that are expressed in languages beyond CQs. A first candidate would be the class of unions of conjunctive queries (UCQs).
2. Extend the setting of Section 4 to more expressive database constraints. Several sufficient conditions for chase termination have been proposed, e.g., *super-weak acyclicity* [17], *c-stratification* [18], *safety* and *inductive restriction* [19], some of which extend stratification and some other are incomparable with it. The question is whether the global constraints satisfy such conditions, as we have shown for the case of stratification when views are defined by conjunctive queries and database constraints are stratified embedded dependencies (and no view constraints).
3. Extend the setting of Section 5 with functional and/or (unary) inclusion dependencies. We conjecture that the former do not influence whether the horizontal decomposition formed by the view symbols is lossless. On the other hand, this is definitely not the case for inclusion dependencies, whose interaction with the domain constraints may entail additional constraints which might influence determinacy, as the following example shows.

   *Example 11.* Consider $\Sigma_{\mathcal{RV}}$ consisting of the following view definition:
   $$\forall x, y \; V(x,y) \leftrightarrow R(x,y) \wedge x > 3 \; ,$$
   and let $\Sigma_\mathcal{R}$ be given by
   $$\forall x, y \; R(x,y) \rightarrow y > 3 \; , \tag{22a}$$
   $$\forall x, y \; R(x,y) \rightarrow \exists z \; R(z,x) \; . \tag{22b}$$
   It is easy to see that $\Sigma_\mathcal{R}$ entails the following additional constraint:
   $$\forall x, y \; . \; R(x,y) \rightarrow x > 3 \; .$$
   Thus, $V$ selects all the tuples in $R$, which is clearly not the case in the absence of the unary inclusion dependency (22b).
4. Study the composition of views (in the sense of functions) and its invertibility. Given that a view obtained by composing two views is invertible if and only if each of the two views in the composition is such, checking for the invertibility of the composition is straightforward whenever one is able to check separately for the invertibility of the views in the composition. The challenge here is understanding how to properly combine two known settings (e.g., those in Sections 4 and 5). Indeed, as the database schema of the second view mapping in the composition is the view schema of the first one, one needs to make sure that certain conditions are satisfied. For example, if $f$ is from $\mathcal{R}$ to $\mathcal{V}$ under $\Sigma$ and $g$ is from $\mathcal{V}$ to $\mathcal{W}$ under $\Gamma$, every $\Sigma$-consistent instance over $\mathcal{V}$ must also be $\Gamma$-consistent. The specific case of combining the settings of Sections 4 and 5 is related to point 3 above, as the resulting scenario with selections on top of views defined by conjunctive queries will most probably require to deal with the interaction between domain constraints and functional dependencies.
5. Consider constraints also on the view schema. In general, it is always possible to allow view constraints such that the $\mathcal{R}$-embedding of the global constraints is a set of constraints for which the solution of the view update problem is already known. For example, in the setting of Section 4, we can add view constraints $\Sigma_\mathcal{V}$ for which the $\mathcal{R}$-embedding $\widetilde{\Sigma}_\mathcal{R}$ of the global constraints $\Sigma$ is a set of stratified embedded dependencies, where $\widetilde{\Sigma}_\mathcal{R}$ is the set of constraints over $\mathcal{R}$ obtained from $\Sigma$ by replacing every occurrence of each *view symbol* with its definition in terms of the database symbols (given in $\Sigma_{\mathcal{RV}}$). What would be interesting to understand is the shape that these view constraints must have in order to satisfy the above condition.

## A Proofs

All the proofs of the general results in Theorems 1–6 can be found in [13]. In the following, we provide the proofs of the results on conjunctive views and horizontal decompositions presented in Sections 4 and 5, respectively.

### A.1 Conjunctive Views

In the technical development of this section, we need to consider instances that, along with constants from **dom**, may possibly contain variables from an arbitrary infinite set **var** of variable names. The active domain of such an instance is the set of constants and variables occurring in it.

A *homomorphism* from an instance $I$ to a instance $J$ is a function $h: \mathbf{adom}(I) \to \mathbf{adom}(J)$ s.t. $h(\overline{a}) \in R^J$ whenever $\overline{a} \in R^I$ and $h(c) = c$ for every constant $c$ occurring in $I$. We write $I \to J$ to indicate that there is a homomorphism from $I$ to $J$. If $I \to J$ and $J \to I$, we say that $I$ and $J$ are *homomorphically equivalent* and we write $I \leftrightarrow J$.

A *universal model* for a set of embedded dependencies $\Sigma$ and an instance $I$ is a finite instance $U$ such that (1) $I \to U$, (2) $U \models \Sigma$, (3) for every (unrestricted) instance $J$, if $J \models \Sigma$ and $I \to J$, then $U \to J$.

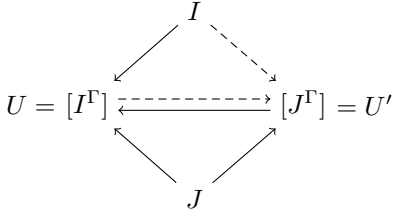

**Figure 3**

We extend the notion of determinacy under constraints to the case in which instances are allowed to contain variables, by considering homomorphic equivalence instead of identity between instances. Thus, we say that $\mathcal{S}$ determines $\mathcal{T}$ under $\Gamma$ w.r.t. homomorphic equivalence, and write $\mathcal{S} \twoheadrightarrow_\Gamma^\leftrightarrow \mathcal{T}$, if for all instances $I$ and $J$ over $\mathcal{S} \cup \mathcal{T}$ such that $I \models \Gamma$ and $J \models \Gamma$, whenever $\mathcal{T}^I \leftrightarrow \mathcal{T}^J$ it is the case that $I \leftrightarrow J$. For ground instances, the two notions coincide, that is, $\mathcal{S} \twoheadrightarrow_\Gamma \mathcal{T}$ iff $\mathcal{S} \twoheadrightarrow_\Gamma^\leftrightarrow \mathcal{T}$, since ground instances are homomorphically equivalent iff they are equal.

For a relation symbol $S$, we denote by $q_S$ the atomic query $S(\overline{x})$, where the number of variables in $\overline{x}$ matches the arity of $S$. Given a query $q$, we define a sentence $\hat{q}$ obtained from $q$ by replacing its free variables with fresh constants.

LEMMA 1. *Let $I = [\{\hat{q}_S\}]$ for some $S \in \mathcal{S}$ and assume there exists a universal model $U = [I^\Gamma]$ for $\Gamma$ and $I$. Let $J = \mathcal{T}^U$ and assume there exists a universal model $U' = [J^\Gamma]$ for $\Gamma$ and $J$. Then, $I \to U'$ whenever $\mathcal{T} \twoheadrightarrow_\Gamma^\leftrightarrow \{S\}$.*

*Proof.* Clearly, $J \to U'$ as $U'$ is a universal model for $\Gamma$ and $J$, and $J \to U$ as $J = \mathcal{T}^U$. Hence, $U' \to U$ by definition of universal model, since $J \to U$ and $U \models \Gamma$. Then, $\mathcal{T}^{U'} \to \mathcal{T}^U$ and, as $\mathcal{T}^U = J \to U'$, also $\mathcal{T}^U \to \mathcal{T}^{U'}$, hence $\mathcal{T}^U \leftrightarrow \mathcal{T}^{U'}$. Since $I \to U$, if $U \to U'$, then trivially $I \to U'$. Conversely, if $I \to U'$, then $U \to U'$ by definition of universal model since $U' \models \Gamma$. Therefore, $U' \leftrightarrow U$ iff $I \to U'$. As $U$ and $U'$ satisfy $\Gamma$ and are such that $\mathcal{T}^U \leftrightarrow \mathcal{T}^{U'}$, assuming $I \not\to U'$ implies $U' \not\leftrightarrow U$ and, in turn, $\mathcal{T} \not\twoheadrightarrow_\Gamma^\leftrightarrow \{S\}$. □

A pictorial sketch of the above proof is given in Figure 3, where the solid arrows denote homomorphisms that exist by assumption while the dashed ones represent homomorphisms derived from the former.

For an instance $I$, let $\varphi_I$ be the sentence obtained by taking the conjunction of all the atomic facts in $I$ and existentially quantifying all the variables. For example, given $I = \{S(a,x), S(y,b), T(x,y)\}$ where $a$ and $b$ are constants and $x$ and $y$ are variables, $\varphi_I$ is $\exists x, y\ S(a,x) \wedge S(y,b) \wedge T(x,y)$.

LEMMA 2. *Every instance $I$ is a universal model for $\{\varphi_I\}$ and $\varnothing$.*

*Proof.* Trivially, $\varnothing \to I$. Moreover, by construction of $\varphi_I$, we have that $I \models \varphi_I$ and, for each symbol $S$ and tuple $\overline{t}$, $S(\overline{t}) \in I$ iff $S(\overline{t})$ is a conjunct in $\varphi_I$. From the latter fact, it easily follows that $I$ has homomorphisms to every instance satisfying $\varphi_I$. □

LEMMA 3. *Let $I$ be an instance and $\Sigma$ a set of embedded dependencies. Then, $[\{\varphi_I\} \cup \Sigma]$ and $[I^\Sigma]$ are homomorphically equivalent, whenever they exist.*

*Proof.* Assume that $U = [\{\varphi_I\} \cup \Sigma]$ and $U' = [I^\Sigma]$ exist. Since $I \to U'$ and $I \models \varphi_I$, we have that $U' \models \varphi_I$, hence $U' \models \{\varphi_I\} \cup \Sigma$. Thus, $U \to U'$ because $U$ is universal for $\{\varphi_I\} \cup \Sigma$ and $\varnothing$. On the other hand, $I \to U$ since $U \models \{\varphi_I\}$ by assumption and $I = [\{\varphi_I\}]$ by Lemma 2. So, $U' \to U$ as $U \models \Sigma$ and $U'$ is universal for $\Sigma$ and $\varnothing$. Hence, $U \leftrightarrow U'$. □

THEOREM 12. *Let $\Gamma$ be a set of constraints over $\mathcal{S} \cup \mathcal{T}$ such that there exists a universal model for every instance. Then, each $T \in \mathcal{T}$ has an exact conjunctive rewriting in terms of $\mathcal{S}$ under $\Gamma$ whenever $\mathcal{S} \twoheadrightarrow_\Gamma^\leftrightarrow \mathcal{T}$.*

*Proof.* Assume $\mathcal{S} \twoheadrightarrow_\Gamma^\leftrightarrow \mathcal{T}$, that is, $\mathcal{S} \twoheadrightarrow_\Gamma^\leftrightarrow \{T\}$ for each $T \in \mathcal{T}$. Let $I$ and $J$ as in Lemma 1 and observe that $U$ and $U'$ exist owing to the main assumption of the theorem (i.e., $\Gamma$ has a universal model for every instance), hence $I \to U'$ by Lemma 1. Let $q$ be the CQ over $\mathcal{S}$ obtained from $\varphi_J$ by substituting all the constants with fresh (free) variables. As $U' \leftrightarrow [\{\varphi_J\} \cup \Gamma]$ by Lemma 3, we have that $q \sqsubseteq_\Gamma q_T$. Since $J = [\{\varphi_J\}]$ by Lemma 2, and $U \leftrightarrow [\{\hat{q}_T\} \cup \Gamma]$ by Lemma 3, from $J \to U'$ we also obtain $q_T \sqsubseteq_\Gamma q$. Therefore, $q_T \equiv_\Gamma q$, that is, $q$ is an exact conjunctive rewriting of $T$ in terms of $\mathcal{S}$ under $\Gamma$. □

Each view definition of the form $\forall \overline{x}\ V(\overline{x}) \leftrightarrow \psi(\overline{x})$, where $V \in \mathcal{V}$ and $\psi(\overline{x})$ is a CQ $\exists \overline{y}\ \varphi(\overline{x}, \overline{y})$ over $\mathcal{R}$, is equivalent to the two following tgd's:

$$\alpha: \quad \forall \overline{x} \qquad V(\overline{x}) \to \exists \overline{y}\ \varphi(\overline{x}, \overline{y}) \qquad (23)$$

$$\beta: \quad \forall \overline{x}, \overline{y}\ \varphi(\overline{x}, \overline{y}) \to V(\overline{x}) \qquad (24)$$

LEMMA 4. *Let $\mathcal{V} = \{V_1, \ldots, V_n\}$ and $\Sigma_{\mathcal{RV}} = \bigcup_{i=1}^n \{\alpha_i, \beta_i\}$, where $\alpha_i$ and $\beta_i$ are of the form (23) and (24), respectively, for each $V_i \in \mathcal{V}$. Then, for every $i, j \in \{1, \ldots, n\}$, it holds that (1) $\alpha_i \not\prec \alpha_j$, (2) $\beta_i \not\prec \beta_j$, and (3) $\beta_i \not\prec \alpha_j$.*

*Proof.* Let $H_i^\alpha$ denote the head of tgd $\alpha_i$ and $B_i^\alpha$ its body, and similarly for $\beta$-tgd's. (1) and (2) follow directly from the fact that $\mathsf{sig}(H_i^\alpha)$ and $\mathsf{sig}(H_i^\beta)$ are disjoint with $\mathsf{sig}(B_j^\alpha)$ and $\mathsf{sig}(B_j^\beta)$, respectively. For an analogous reason, (3) holds whenever $i \neq j$. Thus, it only remain to show that $\beta_i \not\prec \alpha_i$ for every $i$, which follows from the fact that $\forall \overline{x}, \overline{y}\ B_i^\beta(\overline{x}, \overline{y}) \to H_i^\alpha(\overline{x})$. □

COROLLARY 1. *The set of tgd's $\Sigma_{\mathcal{RV}}$ of Lemma 4 is stratified.*

*Proof.* By Lemma 4, the chase graph of $\Sigma_{\mathcal{RV}}$ contains no cycle. □

The following theorem shows that, given $\Sigma_{\mathcal{RV}}$ as above, the global constraints $\Sigma = \Sigma_\mathcal{R} \cup \Sigma_{\mathcal{RV}}$ are stratified whenever the database constraints are such.

LEMMA 5. *Let $\Sigma_\mathcal{R}$ be a set of stratified embedded dependencies over $\mathcal{R}$ and let $\Sigma_{\mathcal{RV}}$ be as in Lemma 4. Then, $\Sigma_\mathcal{R} \cup \Sigma_{\mathcal{RV}}$ is stratified.*

*Proof.* For every $\alpha_i, \beta_j \in \Sigma_{\mathcal{RV}}$ and every $\gamma \in \Sigma_\mathcal{R}$, it is the case that $\beta_j \not\prec \gamma$ and $\gamma \not\prec \alpha_i$, because $\mathsf{sig}(H_j^\beta)$ and $\mathsf{sig}(H^\gamma)$ are disjoint with $\mathsf{sig}(B^\gamma)$ and $\mathsf{sig}(B_i^\alpha)$, respectively. Hence, since the chase graph of $\Sigma_{\mathcal{RV}}$ does not contain any cycle by Lemma 4, cycles in the chase graph of $\Sigma_\mathcal{R} \cup \Sigma_{\mathcal{RV}}$ (if any) consist only of dependencies in $\Sigma_\mathcal{R}$. But $\Sigma_\mathcal{R}$ is stratified by assumption, therefore each such cycle is weakly acyclic. □

We conclude by proving the main results presented in Section 4, namely Theorems 7 and 8.

*Proof (of Theorem 7).*

**"if"** $\mathcal{V} \twoheadrightarrow_\Sigma \mathcal{R}$ whenever there is an exact rewriting of each $R \in \mathcal{R}$ in terms of $\mathcal{V}$ under $\Sigma$, thus in particular when the rewriting is conjunctive.

**"only if"** $\Sigma$ is stratified by Lemma 5, hence there is a universal model for $\Sigma$ and every instance [6]. Since $\mathcal{V} \twoheadrightarrow_\Sigma \mathcal{R}$ implies $\mathcal{V} \twoheadrightarrow^\leftrightarrow_\Sigma \mathcal{R}$, the claim then follows directly from Theorem 12. □

*Proof (of Theorem 8).* Algorithm 1 is a specific instantiation of the C&B, which is sound and complete for finding exact conjunctive rewritings, if any, whenever the chase terminates [7]. In our case, chase termination is ensured by the fact that by Lemma 5 the global constraints we consider are stratified. Algorithm 2 tests, by means of Algorithm 1, for the existence of an exact conjunctive rewriting of each database symbol in terms of the view symbols, and its soundness and completeness follow from Theorem 7. □

## A.2 Horizontal Decompositions

We prove the results of Section 5, namely Theorems 9 and 10, combined together in the following single theorem.

THEOREM 13. *Let $\Sigma$, $\Gamma$ and $\Gamma_{ax}$ be as in Theorems 9 and 10. Then, the following are equivalent:*

1. $\mathcal{V} \twoheadrightarrow_\Sigma \mathcal{R}$;
2. $\Sigma \models \forall \overline{x}\, R(\overline{x}) \leftrightarrow V_1(\overline{x}) \vee \cdots \vee V_n(\overline{x})$;
3. $\Gamma^\alpha$ *is unsatisfiable for every valuation $\alpha$ that satisfies $\Gamma \cup \Gamma_{ax}$.*

In what follows, let **idom** be the domain from which the interpreted attributes take their values (e.g., the integers $\mathbb{Z}$). We first need a technical lemma that will be the key of the main proof.

LEMMA 6. *Let $\alpha$ be a valuation over $\mathsf{var}(\Gamma)$ that satisfies $\Gamma_{ax}$, let $\beta$ be an assignment of values from **idom** to the variables corresponding to the interpreted attributes of $R$, and let $\alpha$ and $\beta$ be such that, for every $v \in \mathsf{cvar}(\Gamma)$,* $\alpha(v) = \mathtt{T}$ *iff* $\beta \models \mathsf{constr}(v)$. *Let $\overline{t}$ be a tuple with the first $k$ values from **dom** and the rest from **idom** such that $\overline{t}[i] = a$ iff $p_i^a \in \mathsf{pvar}(\Gamma)$ and $\alpha(p_i^a) = \mathtt{T}$, for $i \in \{1, \ldots, k\}$, and $\overline{t}[i] = \beta(y_{i-k})$, for $i \in \{k+1, \ldots, |R|\}$. Let $I$ be the instance over $\mathcal{R} \cup \mathcal{V}$ such that $R^I = \{\overline{t}\}$ and $V^I = \varnothing$ for every $V \in \mathcal{V}$. Then, $I \models \Sigma$ if and only if $\alpha$ satisfies $\Gamma \cup \Gamma_{ax}$ and $\beta$ satisfies $\Gamma^\alpha$.*

*Proof.* For any $\psi \in \Sigma$, since by assumption $V^I = \varnothing$ and $R^I = \{\overline{t}\}$, by construction of (10) and (11) $I \not\models \psi$ if and only if all of the following hold:

(a) $\overline{t}[i] = \overline{a}[j]$ $\left(\text{iff } \alpha\big(p_i^{\overline{a}[j]}\big) = \mathtt{T}\right)$ for all $i,j$ s.t. $x_i = \overline{x}'[j]$;

(b) $\overline{t}[i] \neq \overline{b}[j]$ $\left(\text{iff } \alpha\big(p_i^{\overline{b}[j]}\big) = \mathtt{F}\right)$ for all $i,j$ s.t. $x_i = \overline{x}''[j]$;

(c) $\beta \not\models \delta(\overline{y})$ $\left(\text{iff } \alpha(v_\psi) = \mathtt{F} \text{ for } \mathsf{constr}(v_\psi) = \delta(\overline{y})\right)$, if $\psi \in \Gamma_\mathcal{R}$; and

(d) $\beta \models \sigma(\overline{y})$ $\left(\text{iff } \alpha(v_\psi) = \mathtt{T} \text{ for } \mathsf{constr}(v_\psi) = \sigma(\overline{y})\right)$ if $\psi \in \Gamma_{\mathcal{RV}}$.

We show that $I \not\models \Sigma$ iff $\alpha$ does not satisfy $\Gamma \cup \Gamma_{ax}$ or $\beta$ does not satisfy $\Gamma^\alpha$.

**"if"** Assume that $\alpha$ does not satisfy $\Gamma \cup \Gamma_{ax}$ or $\beta$ does not satisfy $\Gamma^\alpha$. We consider the two cases separately.

(1) "$\alpha \not\models \Gamma \cup \Gamma_{ax}$". Then, as $\alpha \models \Gamma_{ax}$ by assumption, there is some propositional formula in $\Gamma$ that is not satisfied by $\alpha$. If the formula is from $\Gamma_\delta$, it has the form $P(\psi)$ as in (10) for some $\psi \in \Sigma_\mathcal{R}$, and $\alpha\big(P(\psi)\big) = \mathtt{F}$ iff (a), (b) and (c) hold, therefore $I \not\models \psi$ and, in turn, $I \not\models \Sigma$. If the formula is instead from $\Gamma_\sigma$, it is of the form $\neg P(\psi)$, with $P(\psi)$ as in (11), for some $\psi \in \Sigma_{\mathcal{RV}}$, and $\alpha\big(\neg P(\psi)\big) = \mathtt{F}$ iff $\alpha\big(P(\psi)\big) = \mathtt{T}$ iff (a), (b) and (d) hold, therefore also in this case $I \not\models \psi$ and, in turn, $I \not\models \Sigma$.

(2) "$\beta \not\models \Gamma^\alpha$". Then, there is a $\mathfrak{C}$-constraint $\phi \in \Gamma^\alpha$ which is not satisfied by $\beta$. Suppose that $\phi \in \Gamma_\sigma^\alpha$, hence $\phi$ is either $\mathsf{constr}(v)$ or $\neg\mathsf{constr}(v)$ for some $v \in \mathsf{pvar}(\Gamma_\sigma)$. By definition of $\Gamma_\sigma^\alpha$, if $\phi = \mathsf{constr}(v)$, then $\alpha(v) = \mathtt{T}$, and in turn $\beta \models \mathsf{constr}(v)$; on the other hand, if $\phi = \mathsf{constr}(v)$, we have that $\alpha(v) = \mathtt{F}$, and in turn $\beta \models \neg\mathsf{constr}(v)$. In both cases, we get a contradiction of $\beta \not\models \phi$, therefore $\phi \in \Gamma_\delta^\alpha$. Then, $\phi = \mathsf{constr}(v) = \delta(\overline{y})$ and, by definition of $\Gamma_\delta^\alpha$, $v$ belongs to a formula $P(\psi) \in \Gamma_\delta$, for some $\psi \in \Gamma_\mathcal{R}$, whose l.h.s. is true under $\alpha$, therefore (a) and (b) hold. Since $\beta \not\models \mathsf{constr}(v)$, we have that $\alpha(v) = \mathtt{F}$, hence also (c) holds. Therefore, $I \not\models \psi$ and, in turn, $I \not\models \Sigma$.

**"only if"** Assume $I \not\models \Sigma$, then there is some $\psi \in \Sigma$ that is not satisfied by $I$. We distinguish the following two cases.

(1) "$\psi \in \Sigma_\mathcal{R}$". Then, $P(\psi)$ has the form (10) and, as (a) and (b) must hold, $\delta(\overline{y}) \in \Gamma_\delta^\alpha$ by definition of $\Gamma_\delta^\alpha$. Therefore, since $\beta \not\models \delta(\overline{y})$ by (c), we have that $\beta \not\models \Gamma_\delta^\alpha$.

(2) "$\psi \in \Sigma_{\mathcal{RV}}$". Then, $P(\psi)$ is of the form (11) and, since (a) and (b) must hold, its truth value under $\alpha$ depends only on the one of the propositional variable $v_\psi$ associated with the $\mathfrak{C}$-constraint $\sigma(\overline{y})$. Indeed, as $\beta \models \sigma(\overline{y})$ by (d), we have that $\alpha(v_\psi) = \mathtt{T}$, thus $\alpha(P(\psi)) = \mathtt{T}$ and, in turn, $\alpha(\neg P(\psi)) = \mathtt{F}$. Therefore, as $\neg P(\psi) \in \Gamma_\sigma$ by definition of $\Gamma_\sigma$, $\alpha \not\models \Gamma_\sigma$. $\square$

*Proof (of Theorem 13).* We show that $3 \implies 2 \implies 1 \implies 3$.

$\boxed{3 \implies 2}$ We prove the contrapositive. Thus, assume $\Sigma \not\models \forall \overline{x}\, R(\overline{x}) \leftrightarrow \bigvee_{i=1}^n V_i(\overline{x})$, that is, there exists a model $I$ of $\Sigma$ such that $R^I \neq \bigcup_{i=1}^n V_i^I$. By definition, $V_i^I \subseteq R^I$ for every $i \in \{1, \ldots, n\}$, and in turn $\bigcup_{i=1}^n V_i^I \subseteq R^I$, hence there must be some tuple $\overline{t} \in R^I$ that does not belong to $\bigcup_{i=1}^n V_i^I$. Let $J$ be the instance such that $R^J = \{\overline{t}\}$ and $V_i^J = \varnothing$ for every $V_i \in \mathcal{V}$. Clearly, $J \models \Sigma$. Therefore, $\alpha$ and $\beta$ as in Lemma 6 satisfy $\Gamma \cup \Gamma_{\text{ax}}$ and $\Gamma^\alpha$, respectively.

$\boxed{2 \implies 1}$ If $R$ has a rewriting in terms of $\mathcal{V}$ under $\Sigma$, then trivially $\mathcal{V} \twoheadrightarrow_\Sigma \{R\}$.

$\boxed{1 \implies 3}$ By contraposition. Let $\alpha$ be a valuation satisfying $\Gamma \cup \Gamma_{\text{ax}}$ and for which $\Gamma^\alpha$ is satisfiable as well, that is, there exists an assignment $\beta$ of values from **idom** to the variables corresponding to the interpreted attributes that makes every formula in $\Gamma^\alpha$ true. As $\alpha$ satisfies $\Gamma_{\text{ax}}$, for every $p_i^a$ and $p_i^b$ in $\mathsf{pvar}(\Gamma)$ with $a \neq b$, it is never the case that both $\alpha(p_i^a) = \mathtt{T}$ and $\alpha(p_i^b) = \mathtt{T}$. Hence, we can find a tuple $\overline{t}$ as in Lemma 6, and an instance $I$ over $\mathcal{R} \cup \mathcal{V}$ s.t. $R^I = \{\overline{t}\}$ and $V^I = \varnothing$ for each $V \in \mathcal{V}$. Since $\beta \models \Gamma_\sigma^\alpha$, by definition of $\Gamma_\sigma^\alpha$ we have that, for every $v \in \mathsf{cvar}(\Gamma_\sigma)$, $\beta \models \mathsf{constr}(v)$ if and only if $\alpha(v) = 1$. Since $\beta \models \Gamma_\delta^\alpha$, by definition of $\Gamma_\delta^\alpha$ we have that, for every $v \in \mathsf{cvar}(\Gamma_\delta)$, $\beta \models \mathsf{constr}(v)$ iff $\alpha(L) = 1$, with $L$ the l.h.s. of the propositional formula in $\Gamma_\delta$ where $v$ appears, and, as $\alpha \models \Gamma_\delta$, in turn $\alpha(L) = 1$ iff $\alpha(v) = \mathtt{T}$. So, $\alpha$ and $\beta$ satisfy the assumptions of Lemma 6, hence $I \models \Sigma$ as $\alpha \models \Gamma \cup \Gamma_{\text{ax}}$ and $\beta \models \Gamma^\alpha$. Then, since $J = \varnothing$ is a model of $\Sigma$ too, and $\mathcal{V}^I = \mathcal{V}^J$ while $R^I \neq R^J$, we conclude that $\mathcal{V} \not\twoheadrightarrow_\Sigma \mathcal{R}$. $\square$